# Smart Bike Sharing System to make the City even Smarter

Monika Rani[*] and O. P. Vyas
Department of Information Technology,
Indian Institute of Information Technology, Allahabad
{monikarani1988 and dropvyas}@gmail.com

**Abstract.** These last years with the growing population in the smart city demands an efficient transportation sharing (bike sharing) system for developing the smart city. The Bike sharing as we know is affordable, easily accessible and reliable mode of transportation. But an efficient bike sharing capable of not only sharing bike also provides information regarding the availability of bike per station, route business, time/day-wise bike schedule. The embedded sensors are able to opportunistically communicate through wireless communication with stations when available, providing real-time data about tours/minutes, speed, effort, rhythm, etc. We have been based on our study analysis data to predict regarding the bike's available at stations, bike schedule, a location of the nearest hub where a bike is available etc., reduce the user time and effort.

## 1 Introduction

Smart city demand for energy-efficient transport system also an emphasis on sharing system for utilization of vehicle (example bicycle, motor bikes etc.). Sharing bike system has various benefits like appropriate resource management, reduce pollution, etc., lead to improved health. To motivate the bike user first we need to explore various parameters like a number of stations, the number of routes, the number of available bikes per station by using the appropriate statistical approach. Once we obtain the correct status regarding the parameters, then the serving of bikes to the growing population in the smart city become quite an easy job.
The motivation of our work is to enhance the efficiency of Bike sharing system by making bikes smart by deploying sensors on bikes this will help in collecting real-time data and to forward them to nearby stations. In the recent years popularity of a bike sharing program has shown a vital growth. Another reason is easily accessible and economical in promoting short-term bike rental system. There would be less congestion for commuters in traffic of the car parking. Tourists can also enjoy hassle free travel without changing multiple busses and taxis. The environment also gets the benefit of less smog after a weekday commute. Cities are managing development and urban living culture is facing major challenges in our daily lives. Based on statistical data of 2007, half of the population of the world was living their lives in cities. UN population fund forecasts that by the end of 2030 nearly 60% of the world population would live their life in cities [1]. Out of all major issues, we can outline, air quality, environmental crisis, and transportation issues. Use of a bicycle is an important mode of transportation that could be helpful to many urban transportation issues. As the use of motor vehicles is increasing problems as cost, congestion, accidents, loss of amenity and space, noise, air pollution, energy consumption and have an adverse effect on the natural environment. In future, use of the bicycle as transport should be the transportation solution for cities as it has no adverse effects [2] [3].

One more critical issue in any modes of transportation is of pod cars to bike, the issue is about picking people from the transport hub as a railway station, bus depots and to their destination, this problem is called last mile problem [4]. Ever wondered if we could take our vehicle out and then forget it after reaching the destination or if you don't have to plea anyone to drop you down or pick you up. Bike sharing is required where a person can borrow a bicycle from one of our stations in the city and can return to another station. For the same smart bike sharing system develops the concept of riding a bike from one point and returning it back at another point can help to solve last mile problem. An efficient bike sharing system collects the real-time data using sensors deployed on them and send the data to stations on the way. Smart bike sharing not only solves the issue of the last mile, but also a problem of area/acres requires for car parking and reduces the waiting time for a local bus.

Our problem was to increase the efficiency of bike sharing systems and increase the participation of people in these sustainable systems, thus moving towards the concept of the smart city. Due to growing population and change in the transportation usage in urban cities, some people have demanded usage of bike sharing system in the last few years [5] [6], however, still there are some people who are the reluctance to a combination of the mode of transportation of traditional days [7]. Some explanation was given by people for choosing a traditional transportation, over bike sharing system is that the bike riding is not safe and one has to travel a longer distance also, the weather is another deciding factor. The transportation authorities and local council need to encourage the use of a bike and also need do develop separate infrastructure and alternative routes for bikes to provide safety and provide shorter distance [8] [9]. Also, it might be the case that at any point of time the sharing station can be empty and there is no bike to rent on the other hand it also be possible that people unable to park the bikes due to the station is full and there is no availability of parking slots. The problem can be extended to use other public transport systems so that whole city transport can be transformed to make the city, even smarter. These problems gave us the motivation to innovate and implement a bike sharing system that is different and better than existing systems.

The paper is structured into 5 sections; initially, we focus on the introduction of the paper in section 1. In Section 2, we describe related works, where we have explained the works that have been done on a bike sharing system and proposed in the relative fields of sensors and smart cities. Section 3 describes a website to enable users of the system to book the bikes from any station, see the station map and check the current availability and see their previous rides. Virtually demonstrating the system using NS2 Simulator and collecting the simulation data along the way. The section also depicts experimental results that prove the usefulness of the smart bike sharing system. Basically, our analysis of data from a similar existing system for minimizing the redistribution problem and proposing such analysis when actual data is collected in our system. Section 4 discussion focus on contribution of work and limitations of work. Finally, section 5 draws the conclusion and future research opportunities in the bike sharing system for smart cities.

## 2  Literature review of related work

Much has been said about how making a city smart and many international conferences have taken place on this issue for a sustainable future. Many researchers have published their research on using a bike sharing system to make the city smarter and how to analyze the bicycle sharing data for generating insights into sustainable transport systems. We have primarily taken guidance from one research paper which deals with how the bikes will send the data on the way and what will be the protocol that will govern this transmission [10] and also took some motivation for better analysis from another research paper which has discussed an analysis of various bike share systems currently existing in various countries [11]. The bike sharing system can use various technologies to make it smart and real time example, advances sensors, collaborative agents and ontologies for storing user's and heterogeneous station details [12] [13]. The former research paper has thoroughly covered the topic that was needed at work and some of the terms use like bike availability, busiest station, and load factor etc., and also characteristics of docking stations describe as follows: Aggregate characteristics, Spatial characteristics, Temporal characteristics, Demographic and community detection of data and the redistribution problem.

Several cities in the United States and Europe have started implementing bike sharing programs in their cities, a few of which we encountered during our research on the project are: Germany has bike-sharing programs in many cities, including Aachen, Berlin, Cologne, Düsseldorf, Frankfurt, Hamburg (StadtRAD Hamburg), Karlsruhe, Kassel (Konrad), Mainz (MVGmeinRad), Munich and Stuttgart. Station based System Metropolradruhr is located in the Ruhr Area. Bike-sharing stations are also located in over 50 ICE railway stations [14] [15]. French cities offering a sharing system include Marseille, Lyon, Bordeaux, Nice, Toulouse, Rennes, Rouen, La Rochelle, Orléans, Montpellier, Nantes, Lille, Strasbourg, Clermont-Ferrand, Avignon, Saint-Étienne, ChalonsurSaône, Belfast, and Aix-en-Provence. Similar work has been done in cities like California and San Francisco [16] [17]. For the simulation, we referenced the standard tutorial for NS2 by MarcGreis [18].

## 3   Methodology

A website to enable users of the system to book the bikes from any station, see the station map and check the current availability and see their previous rides. For the same a workflow of a bike sharing system for users is shown in Fig. 1 which is capable of keeping detail regarding reserve bikes, see their previous rides (as shown in Fig. 2) and get to know about how bike sharing works. The deployed sensor on the bikes which is used to collect the data this data can be further used for predictive user requirement like a number of bikes available at a particular station, route business, the average time is taken between two stations [19]. Simulating the smart bike share system with the help of NS2 Simulator and plotting the graphs of data send by bikes and data received at a station as shown in Fig. 3. Analysis of data from a similar existing system for minimizing the redistribution problem and proposing such analysis when actual data is collected in our system. Equipping bikes with sensors to retrieve the information to draw an inference for various aspects like air pollution, noise pollution, and other effects on human and its lifestyle.

**3.1 Data analysis**

There is the following analysis that has done on available data in our paper:
The description of the data set (www.kaggle.com) [20] which consisted of following files and further described are attributes of each file.
 i. 201402_status_data.csv– approx. 17 million records of Status Data (Bikes_available, Dock_available, and Time)
 ii. 201402_station_data.csv – 69 records – Station Data (Station_ID, Name, Latitude, Longitude, Dock_count, Landmark, and Installation)
 iii. 201402_trip_data.csv – approx. 144,000 records of individual trips. Trip Data (Trip_ID, Duration, Start _Date, Start_ Station, Start_Terminal, and End_Date, End_Station, End_Terminal, Bike_No., Zip_Code, Subscription_Type).

- Prediction of bike availability

We have analyzed our available data to get the prediction that what will be the condition of bike availability in the future at a particular station, this analysis will help to inform the users that at a particular time bike will be available or not on that station also informs users about the estimated waiting time for a bike, example as a number of bikes that should be available at the station shown in Fig. 4.
We have used the concept of weighted arithmetic mean [21] on some parameters that are listed below (Suppose today's date is Wednesday, $18^{th}$ November 2016):
 i. No. of bikes at a Day of the Week (DoW) (25% weight) – We analyzed previous data for last 6 Wednesdays.
 ii. No. of bikes in Current Week (CW) (50% weight) – Analyzing the trends of the current week and giving them a weight of 50%.
 iii. No. of bikes on the Day of the Month (DoM) (25 % weight) – Analyzing the data for the current date of last 6 months (last 6 18's in the current year).

No of bikes that should be available today at stations X:

$$N = 0.25 * DoW + 0.5*CW + 0.25*DoM$$

- Busiest station

We have analyzed the data to find the busiest stations (busiest means where bike incoming and bike outgoing is more) as shown in Fig. 5. Analysis of data of last 6 months revealed the stations that are most busy. For a station X busyness is denoted by K. And the total no. of incoming bikes in a station X (Incoming data = L) & total no. of outgoing bikes from station X (Output data = M). For a station X, the busyness is calculated as:

$$K = L + M$$

The busiest station at a particular time or day. At a particular time, which station has more rush, this information will help the user to know about the availability of bike at a given time. We have analyzed data for last 6 months and calculated density of busyness of all stations at a particular hour. The busyness of a station X at the hour t is defined as the sum of no. of all incoming + outgoing bikes from station X in the interval t to t + 1. Example our approach tries to provide useful information like the number of bike trips from each station and the density of busyness of stations at a particular time as shown in Fig. 6 and Fig. 7 respectively. Also, we can calculate which station has a maximum number of incoming bikes at a particular time.

- The Average time took between two stations

The average time is taken by the user from a station to reach other stations. We analyzed trip data and took the arithmetic mean of all the trip times for trips between X to Y ($xy_1$, $xy_2$, $xy_3$...$xy_n$) and Y to X ($yx_1$, $yx_2$, $yx_3$...$yx_m$) in the past 6 months, which gave us an average trip time between two stations. For calculating trip times between stations X and Y:

$$((xy_1, xy_2, xy_3...xy_n) + (yx_1, yx_2, yx_3...yx_m))/N+M$$

- The Busyness of all routes

To analyze the busyness of all routes, we analyzed trip data between two stations. This information helps the user to reduce their trip time. The busyness of a route between two stations is defined as the total number of trips between those two stations over the past 6 months.

- Load Factor

The load factor of a particular station as shown in Fig. 8. The calculated load factor (P) of a station represents the load a station is bearing. Load factor is calculated by the sum of a number of available bikes and a number of empty docks. Whereas, number of available bikes and number of empty docks are denoted by Q and R respectively. load factor increases are directly propositional to load increases on a station. The load factor for a station is calculated as:

$$P = Q + R$$

- Estimated time of availability of a bike

We try to calculate the probability of availability of a bike for each of the next 30 minutes as shown in Fig. 9. The data set (status_data describe attributes Bikes_available, Dock_available, and Time). The attribute time is used to calculate the probability of availability of bike. The attribute time is actually timestamp of the form (YYYY-MM-DD HH: MM: SS). Let Max_bikes (U) denote the maximum number of bikes that can be present at a station X for which the probability is to be calculated. Suppose today's date is 1st June, 2016 (Wednesday) and the time at which a user arrives at a station and finds zero bikes available is 15:45:00. In our bike sharing system, we attempt to predict and recommend to the user whether he/she should wait for an incoming bike or not. The prediction is based on calculating the probability for an availability of a bike for the next 30 minutes. To calculate the probability for an availability of a bike for the next 30 minutes we account three factors of the current timestamp (at which user finds zero bikes available).

  i. No. of bikes on the day of a week (Wednesday) at the current time (t) (for the last 6 Wednesdays) denoted by V1.
 ii. No. of bikes in the current week at the current time (t) - this includes all the days of the current week denoted by V2.
iii. No. of bikes on the day of a month (1st of every month) for the last 6 months at the current time (t) denoted by V3.

Let t denote the time at which the probability is to be calculated, t ranges from (time of user arrival) to (time of user arrival + 30 minutes). Let w1, w2 and w3 denote the weights assigned to the factors (V1, V2, and V3) experimentally. Therefore, probability of availability of a bike at t'th minute (T) can be calculated by:

$$T = (w1 * V1 + w2 * V2 + w3 * V3)/U$$

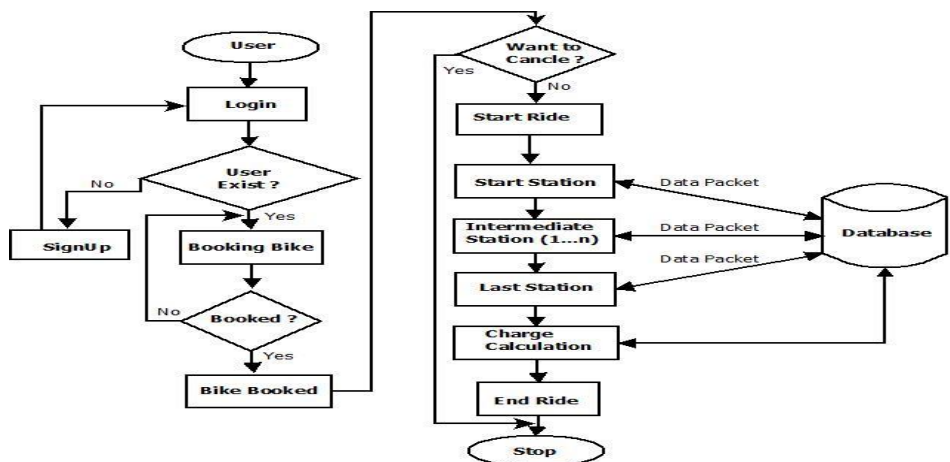

Fig. 1. Workflow of smart bike sharing system

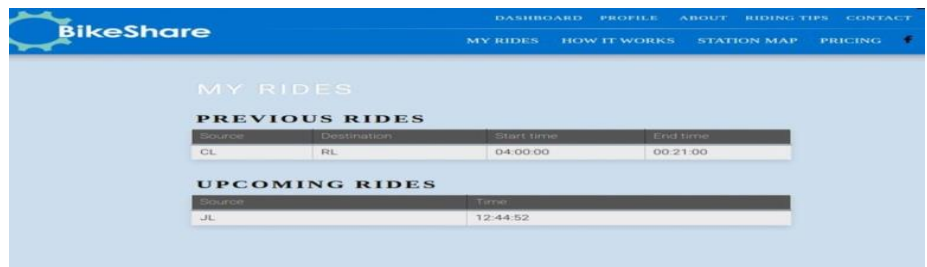

Fig. 2. User's previous and upcoming rides

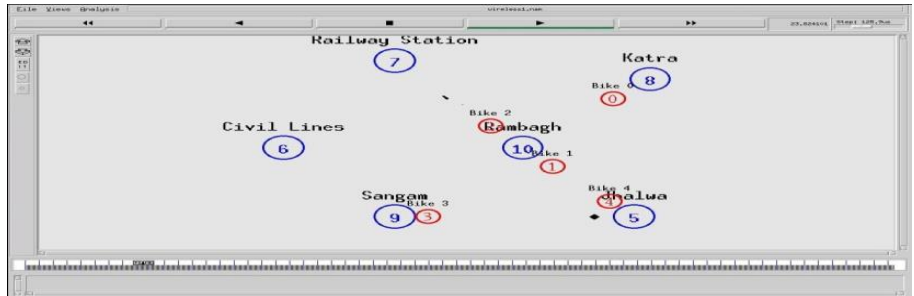

Fig. 3. NS2 Simulating the smart bike share system

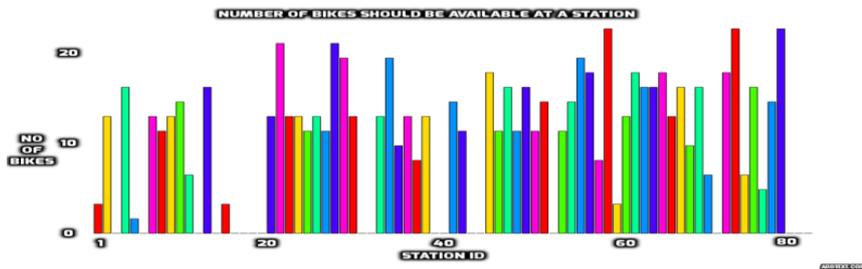

Fig. 4. Number of bikes that should be available at the station tomorrow

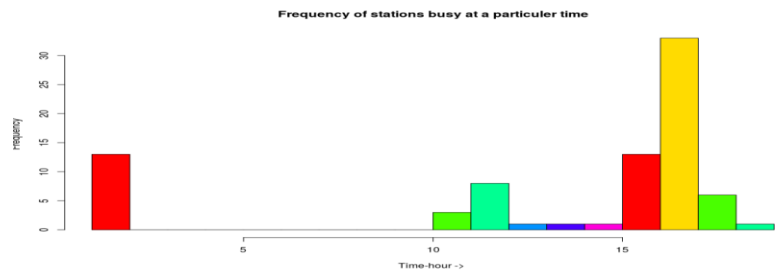

Fig. 5. Number of stations that are busy at a particular time

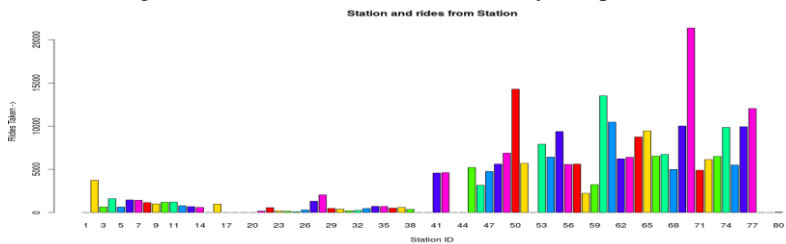

Fig. 6. Number of bike trips from each station

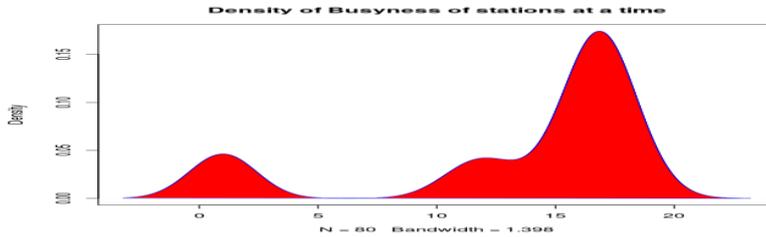

Fig. 7. Density of busyness of stations at a particular time

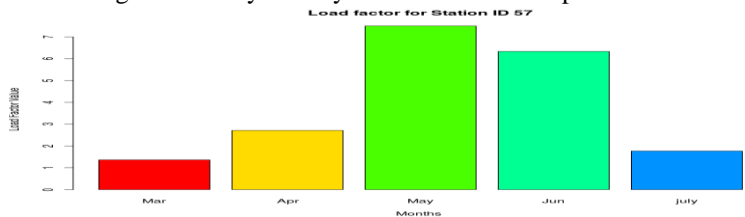

Fig. 8. Load factor of a particular station

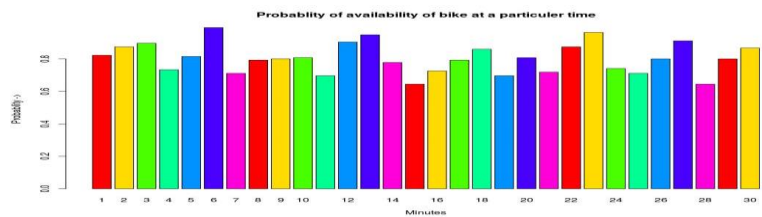

Fig. 9. Probabilities of availability of a bike for each of next 30 minutes

**3.2 Simulating the smart bike sharing system using the NS2 Simulator**

Better understanding of bicycle habits, path, utilization rate. Making bikes intelligent by deploying sensors that send real-time data to base stations. We have simulated the system with the help of network simulator. Key features of simulation are:

- Bike starts from the source and ends at, a destination, both of which are not known a priori. We have tried to show that our system will work under any random scenario which may occur in practice. Simulator starts n (taken as command line argument) bikes from 6 previously defined stations and marks the destinations of the bikes also (randomly). Then each bike starts a trip at a random time (which is the case that will occur in practice).
- On the way to the destination, a bike collects data via sensors and sends the data to stations following AODV (Ad-hoc on demand Distance Vector routing) protocol [22].
- When a bike comes in the range of a station, it makes a TCP connection with the station and sends the data to it.
- If a bike comes in the range of multiple stations, it sends equal amounts of data to all of them, thus leading to more network utilization. After simulation, we get the following data in the trace file:
    i. The Amount of data sent by each bike.

ii. The Amount of data received by each station.

## 4 Discussion

### 4.1 Contribution of work

We have tried to give extended functionality and eliminate the flaws of some of the existing bike share systems. The facilities provided by our system are:
i. Bikes can be booked sitting at home using our website so that users do not have to wait in case of unavailability.
ii. Previous years' data have analyzed to avoid unavailability.
iii. Hubs will be scattered densely all over the city. Users can take a bike from any hub and return it to any hub in the city.
iv. The whole system will be automatic, there will be no need of human assistance on each hub.
v. Sensors will be deployed in bikes that will collect data on the way and send to the nearest station.

### 4.2 Limitations of Work

i. The current system does not identify the type of data sent by the bike.
ii. Continuously updating the data received from bikes in the database. And analyzing the data set of about over 17 million records. Also, the availability of data with respect to a particular country is a challenge.
iii. When the bike is not in range of any station, it will not be able to send the data and we will try to use the protocol required for multi-hop communication.

## 5 Conclusion and Future work

In this paper, we have represented detailed design, implementation plan and evaluation of smart bike sharing system along with sensor networking techniques. The bike sharing system represents the first comprehensive mobile sensing system conveying the cyclist experience. Bike sharing provides the collection and communal environmental sampling. It also supports two modes of operation in support of delay tolerant and real-time sensing. Collected data could be presented both locally to the cyclist and to others as well through back-end services. Bike sharing portal concept promotes social and friendly network among cyclists. Our smart bike sharing system allows the users to easily book a bike using the website at any time without human intervention. There is no need of human for conducting this smart bike sharing system. A user can take a bike from the station using his/her smart card (a smart card that will be given to the user after the SignUp) and start the ride and after completing the trip drop the bike to the station which is near to his/her destination. The Simulator in our system is using the sensor to trace the bike and to update the information of the bike position at each time. The sensor will send packets to its nearest station and these all station will be connected to the website and send the information regarding the bike's status to the app which will update the record.

The data send by the bike can include Traffic data, Air Quality data, Road Conditions data etc., which will benefit the operator in solving the redistribution problem as well as user's of the system thus saving on operational cost as well as the time of users. In the future when conventional sources of energy would be scarce, bike share system will provide an effective means of transport and within the city, it can be made compulsory to travel through bicycles. In future, our smart bike sharing system can be improved by using collaborative software agents on user's and station details store on ontologies. Ontologies can easily expand with the addition of users and stations in the system, provide a secure environment, and machine-readable data for agent's interaction. Software agents can monitor data packets at heterogeneous stations to provide real-time information.